\documentclass[%
aps,
reprint,
superscriptaddress,
noeprint,
amsmath,amssymb,
prb,
floatfix,nofootinbib
]{revtex4-2}

\usepackage{chemformula}
\usepackage[utf8]{inputenc}
\usepackage[T1]{fontenc}
\usepackage{graphicx}
\usepackage{epstopdf}
\usepackage{siunitx}
\usepackage{amsmath}
\usepackage{amssymb}
\usepackage[colorlinks=true,allcolors=blue]{hyperref}
\usepackage{soul}
\usepackage[normalem]{ulem}
\usepackage{lipsum}
\usepackage{xcolor}
\usepackage{pdfpages}
\usepackage{pgffor}

\makeatletter
\AtBeginDocument{\let\LS@rot\@undefined}
\makeatother

\newcommand{\orcid}[1]{\href{https://orcid.org/#1}{\includegraphics[width=8pt]{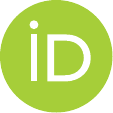}}}

\begin{document}

\title{Amorphous Bi$_2$Se$_3$ structural, electronic, and topological nature from first-principles}

\author{Bruno Focassio\orcid{0000-0003-4811-7729}}\email{b.focassio@ufabc.edu.br}
\affiliation{Federal University of ABC (UFABC), 09210-580 Santo André , São Paulo, Brazil}
\affiliation{Brazilian Nanotechnology National Laboratory (LNNano), CNPEM, 13083-970 Campinas, São Paulo, Brazil}

\author{Gabriel R. Schleder\orcid{0000-0003-3129-8682}}
\affiliation{Federal University of ABC (UFABC), 09210-580 Santo André , São Paulo, Brazil}
\affiliation{Brazilian Nanotechnology National Laboratory (LNNano), CNPEM, 13083-970 Campinas, São Paulo, Brazil}
\affiliation{John A. Paulson School of Engineering and Applied Sciences, Harvard University, Cambridge, Massachusetts 02138, USA}

\author{Felipe Crasto de Lima\orcid{0000-0002-2937-2620}}
\affiliation{Brazilian Nanotechnology National Laboratory (LNNano), CNPEM, 13083-970 Campinas, São Paulo, Brazil}
\affiliation{Ilum School of Science, CNPEM, 13083-970 Campinas, São Paulo, Brazil}

\author{Caio Lewenkopf\orcid{0000-0002-2053-2798}}
\affiliation{Instituto de F\'{\i}sica, Universidade Federal Fluminense, 24210-346 Niterói, Rio de Janeiro, Brazil}

\author{Adalberto Fazzio\orcid{0000-0001-5384-7676}}\email{adalberto.fazzio@lnnano.cnpem.br}
\affiliation{Brazilian Nanotechnology National Laboratory (LNNano), CNPEM, 13083-970 Campinas, São Paulo, Brazil}
\affiliation{Federal University of ABC (UFABC), 09210-580 Santo André , São Paulo, Brazil}
\affiliation{Ilum School of Science, CNPEM, 13083-970 Campinas, São Paulo, Brazil}

\date{\today}

\begin{abstract}
Crystalline $\rm Bi_2Se_3$ is one of the most explored three-dimensional topological insulator, with a $0.3\;\rm eV$ energy gap making it promising for applications. Its amorphous counterpart could bring to light new possibilities for large scale synthesis and applications. Using \textit{ab initio} molecular dynamics simulations, we have studied realistic amorphous $\rm Bi_2Se_3$ phases generated by different processes of melting, quenching, and annealing. Extensive structural and electronic characterizations show that the melting process induces an energy gap decrease ruled by growth of the defective local environments. This behavior dictates a weak stability of the topological phase to disorder, characterized by the spin Bott index. Interestingly, we identify the occurrence of topologically trivial surface states in amorphous $\rm Bi_2Se_3$ that show a strong resemblance with standard helical topological states. Our results and methods advance the search of topological phases in three-dimensional amorphous solids.
\end{abstract}

\maketitle 

\section{Introduction}

Topological insulators (TIs) are quantum states of matter with an insulating energy gap in the bulk and metallic boundary states that are robust against the presence of disorder \cite{Hasan2010,Ando2013a,Bansil2016}.
Topological phases of matter in both two  (2D) and three-dimensional (3D) systems gained significant interest due to their exotic properties and potential applications \cite{Moore2010a, Qi2011, Zhang2019, Vergniory2019, Giustino2021}.

Topological properties of 3D materials have been predicted and observed in ${\rm Bi}_{1-x}{\rm Sb}_x$ alloys \cite{Fu2007,Teo2008a,Hsieh2008}, and the family of layered chalcogenides $\rm Bi_2Se_3$, $\rm Bi_2Te_3$, and $\rm Sb_2Te_3$ materials \cite{Zhang2009, Xia2009, Hsieh2009, Chen2009}. 
Layered Bi and Sb chalcogenides have been extensively studied due to their enhanced thermoelectric properties \cite{Ghaemi2010, Hinsche2011, Chang2014, Liu2017, Yu2014}. Several works have demonstrates a variety of interesting electronic properties of the topological phases in these materials: doping to control of the electronic structure and surface states \cite{Abdalla2013, Jurczyszyn2020, Holtgrewe2021, Ptok2021}; alloying induced topological phase transition \cite{Liu2013}; temperature effects in the electronic structure and temperature-induced phase transition \cite{Monserrat2016,Antonius2016}; spin-polarized transport and spin-orbit torque \cite{Yazyev2010a, Costache2014, King2011}; and unique properties of heterostructures \cite{Seixas2013, Freitas2016, DeOliveira2017, Costa2018, Wang2015}. Additionally, compounds belonging to the class of chalcogenides alloys are candidates for optical data storage, and electronic phase change memories due to a fast transition between crystalline and amorphous phases \cite{Wuttig2007, Lacaita2008}. For instance, detailed theoretical investigation of the structure of amorphous $\rm GeTe$, $\rm Ge_2Sb_2Te_5$ (GST), $\rm Sb_2Te_3$, $\rm Bi_2Te_3$, and $\rm In_3SeTe_2$ relate the speed of this transition to the volume of nanocavities and defective local environments enhancing diffusivity \cite{Caravati2007, Caravati2010, PCCPguo2018, Guo2019, Los2013}.

Recent works investigating 2D and 3D model Hamiltonians have shown that non-crystalline and amorphous systems can also support topological phases \cite{Agarwala2017a, Yang2019a, Mukati2020, Wang2020}. These findings have the potential to vastly expand the field, provided one finds material system realizations.
  
In 2D, amorphous bismuthene has been predicted to display a robust quantum spin Hall topological phase \cite{Costa2019a, Focassio2021}.
In 3D, the scenario is less clear. Amorphous phases of the rhombohedric A$_2$X$_3$ materials, e.g., Sb$_2$Te$_3$ \cite{PCCPguo2018}, Bi$_2$Te$_3$ \cite{Guo2019}, Bi$_2$Se$_3$ \cite{JAPbarzola2015}, have been experimentally achieved. In amorphous $\rm Bi_2Se_3$, experimental evidence suggests a surface Dirac cone with helical spin texture \cite{Corbae2019}. In Sb$_2$Te$_3$, the topological phase vanishes with disorder, and spin correlations dominate the charge transport \cite{NPJQMkorzhovska2020}. However, the topological nature of amorphous $\rm Bi_2Se_3$ family of materials has not been theoretically addressed, and even less so a topological-trivial phase transition driven by the structural phase transition between the crystalline and the amorphous phases in these materials.

In this work, we use \textit{ab initio} molecular dynamics to generate realistic amorphous $\rm Bi_2Se_3$ ($a$-$\rm Bi_2Se_3$) systems and study their structural and electronic properties, as well as their topological classification. We assess the coordination and quality of the local environments of the Bi and Se atoms and observe a growing number of defective environments with increasing temperature. Our main finding is that the topological properties of $\rm Bi_2Se_3$ are not robust against structural disorder: The growth of defective octahedral environments drives a topological phase transition. More specifically, our simulations indicate that the crystalline-amorphous transition at $ T \approx 1,600$ K is accompanied by a topological-trivial phase transition. 
We also show that near the transition point, disordered $\rm Bi_2Se_3$ structures have surface states that resemble the helical Dirac cone of topological insulators.

This paper is organized as follows. In Sec. \ref{sec:methods} we present the computational methods used in this study for the molecular dynamics and electronic structure calculations. We also propose a 3D generalization of the  standard procedure to compute the spin Bott in 2D systems. In Sec.~\ref{sec:results} we discuss the structural, electronic, and topological properties of amorphous $\rm Bi_2Se_3$. This analysis serves as a basis to study the structural and topological phase transition of the material as a function of the temperature. We conclude by  summarizing our findings and presenting an outlook in Sec.~\ref{sec:conclusions}.

\section{Computational methods}
\label{sec:methods}

We generate the amorphous structures with \textit{ab initio} molecular dynamics calculations (AIMD) using Density Functional Theory (DFT) \cite{Hohenberg1964, Kohn1965, Schleder2019} as implemented in the Vienna \textit{Ab Initio} Simulation Package (VASP) \cite{Kresse1996, Kresse1996c}. The structures are generated by the usual melt-and-quench procedure. 
Starting from the crystalline structure with a 135 atoms supercell, we heat the system up to $2000\;\rm K$. After annealing at $2000\;\rm K$ for $10\;\rm ps$, we quench the system from $2000\;\rm K$ to $300\;\rm K$ in $10\;\rm ps$ and then anneal it once more at $300\;\rm K$ for another $10\;\rm ps$. In the AIMD simulations, we use $2\;\rm fs$ as the time-step for integrating the equations of motion using the Nosé-Hoover thermostat \cite{Nose1984, Nose1984b, Hoover1985} for generating the NVT ensemble. After the final annealing, we perform a structure and variable cell relaxation until the Hellman-Feynman \cite{Feynman1939} forces are less than $2.5 \times 10^{-2}\;\rm eV \si{\angstrom}^{-1}$. The electronic exchange and correlation interaction is treated by the generalized gradient approximation (GGA) \cite{Perdew1992} with the Perdew-Burke-Ernzerhof (PBE) \cite{Perdew1996a} exchange and correlation functional. We use the projector augmented wave (PAW) \cite{Kresse1999a} for describing the ionic cores. The calculations are performed with a $300\;\rm eV$ kinetic energy cutoff for the plane-wave expansion and using only the $\Gamma$-point for sampling the Brillouin zone (BZ). 
Density of states are calculated with a $3\times 3\times 1$ $\Gamma$-centered Monkhorst-Pack k-point grid. The spin-orbit coupling (SOC) interaction was included in all density of states and eigenvalues calculations. 
In Sec. \ref{sec:results} we discuss the results for a representative structure obtained with this procedure. The results for similar amorphous structure obtained with NVT and also NPT ensemble can be found in the Supplemental Material \cite{SuppMat}.

To deepen our understanding on the topological nature of the amorphous structures, we extract a local basis Hamiltonian based on pseudo-atomic orbitals (PAOs) as implemented in the PAOFLOW code \cite{Agapito2013, Agapito2016, Agapito2016a, BuongiornoNardelli2018}. The PAO Hamiltonian is built from the projection of the Kohn-Sham (KS) Bloch states onto the basis of Bi-\textit{spd} orbitals and Se-\textit{sp} orbitals from the pseudopotentials. The effective Hamiltonian reads

\begin{equation}
    H = \sum_{ij}\sum_{\mu\nu}\sum_{\sigma} t_{ij}^{\mu\nu} c_{i\mu\sigma}^{\dagger} c_{j\nu\sigma} \;\text{,}\label{eq:ham_paoflow}
\end{equation}
where the operator $c_{i\mu\sigma}^{\dagger}$ ($c_{j\nu\sigma}$) creates (annihilates) an electron with spin projection $\sigma$ at the atomic site $i$ ($j$) and orbital $\mu$ ($\nu$). The hopping matrix elements $t_{ij}^{\mu\nu}$ are obtained directly from the projection of KS states onto PAO orbitals without parameter fitting. The KS states are obtained from DFT calculations based on the Quantum Espresso (QE) \cite{Giannozzi2009, Giannozzi2017} code with PBE exchange and correlation functional, $80\;\rm Ry$ for the wavefunction kinetic energy cutoff. For computational efficiency, we exclude the Bi-\textit{d} orbitals and include SOC on the PAO Hamiltonian via an effective approximation \cite{Abate1965}, namely
\begin{equation}
    H_{SOC} = \sum_i \sum_{\mu\nu} \sum_{\sigma\sigma^\prime} \lambda_i\langle i \mu \sigma | \mathbf{L} \cdot \mathbf{S} | i \nu \sigma^\prime \rangle c_{i\mu\sigma}^{\dagger} c_{j\nu\sigma^\prime} \;\text{,} \label{eq:ham_soc_paoflow}
\end{equation}
where the $\mathbf{L}$ and $\mathbf{S}$ are the orbital and spin angular momentum operators. The SOC parameter $\lambda$ for the Bi-\textit{p} and Se-\textit{p} orbitals are $\lambda_{\rm Bi} = 1.615\;\rm eV$ and $\lambda_{\rm Se} = 0.320\;\rm eV$, which accurately reproduces the bulk band structure with self-consistent SOC, see Ref. \cite{SuppMat}.

The classification of topological insulators without translational symmetry, like amorphous systems, requires a real-space invariant since the reciprocal-space Chern number is undefined. In this paper we use the spin Bott index, a standard invariant of this kind. 
The Bott index indicates obstructions to deform the wave functions of the filled states of a given system into completely localized orbitals, which is equivalent to reciprocal-space invariants in the infinite-system limit \cite{Hastings2010}. 
We use the PAO Hamiltonians to implement and compute the spin Bott index \cite{Loring2010, Hastings2010, Huang2018, Huang2018a}. 

For that purpose, we put forward a three-dimensional generalization of the standard method to compute the Bott index, otherwise restricted to address 2D systems. For the sake of clarity, let us begin by briefly reviewing how the spin Bott index is calculated. First, one constructs the projection operator of the occupied states,
\begin{equation}
    P = \sum_{n}^{N_{\rm occ}} |\psi_n \rangle\langle \psi_n| \;\text{.}\label{eq:proj_occ}
\end{equation}

Next, one decomposes the occupied space projector into spin-up ($+$) and spin-down ($-$) sectors by constructing the spin operator projected onto the occupied states, namely
\begin{equation}
    P_z = P \hat{s}_z P  \;\text{,}\label{eq:proj_Pz}
\end{equation}
where $\hat{s}_z = \frac{\hbar}{2} \sigma_z$ is the spin operator expressed, as standard, in terms of the Pauli matrix $\sigma_z$. The projector of the occupied spin-up ($+$) and spin-down ($-$) sectors are constructed from the eigenstates of $P_z$ as
\begin{equation}
    P_{\pm} = \sum_{n}^{N_{\rm occ}/2} |\phi_n^{\pm} \rangle\langle \phi_n^{\pm}| \;\text{,}\label{eq:proj_occ_sector_Ppm}
\end{equation}
 where $|\phi_n^{\pm}\rangle$ are the eigenstates of $P_z$ with eigenvalues $S_{\pm}$. 
In the case of 2D systems, for each spin sector one then constructs the projected position operators
\begin{align}
    U_{\pm} = P_{\pm} e^{i2\pi X} P_{\pm} + (I - P_{\pm}) \;\text{,}\label{eq:U_proj_position_operators} \\
    V_{\pm} = P_{\pm} e^{i2\pi Y} P_{\pm} + (I - P_{\pm}) \;\text{,}\label{eq:V_proj_position_operators}
\end{align}
where $X$ and $Y$ are diagonal matrices with the $x$ and $y$ components of the spatial coordinate of each orbital site rescaled to unity. 
To improve the numerical stability of the spin Bott index \cite{Huang2018, Huang2018a} it is useful to use the singular value decomposition (SVD) for the projected position operators $U_{\pm}$, $V_{\pm}$.
The Bott index of each spin sector is defined as \cite{Bellissard1994, Exel1991, Loring2010, Hastings2010, Huang2018, Huang2018a}
\begin{equation}
    B_{\pm} = \frac{1}{2\pi} {\rm Im} \left\{ {\rm Tr} \left[ \log \left( V_{\pm}U_{\pm}V_{\pm}^{\dagger}U_{\pm}^{\dagger} \right) \right] \right\} \;\text{.}\label{eq:BottIndexSector_Bpm}
\end{equation}

Finally the spin Bott index is defined as
\begin{equation}
    B_{s} = \frac{1}{2} (B_{+} - B_{-}) \;\text{,}\label{eq:spinBott_Bpm}
\end{equation}
similar to the spin Chern number \cite{Sheng2006, Prodan2009, Prodan2010, Prodan2011, Sheng2013}. It is important to stress that the error of the the (spin) Bott index scales with with system size as $L^{-1}$ \cite{Toniolo2017}. This means that one needs to consider large supercells in order to obtain a converged calculation of the topological invariant. For our amorphous systems, obtained using large simulation cells with accurate PAO Hamiltonians, this can be computationally demanding. Additionally, we note that the formulation of the spin Bott index is valid in the presence of spin-mixing terms such as strong SOC and Rashba terms, as long as there is a gap in the spectra of the eigenvalues of $P_z$, {\it i.e.}, the eigenvalues are close to the eigenvalues of the spin matrix $\hat{s}_z$, $\pm\hbar/2$ and zero, and exists a gap between those \cite{Huang2018a}, which is our case for amorphous insulators.

The above formulation of the spin Bott index is defined for 2D systems. For 3D systems we propose the use of a spin Bott index vector, similar to the $\mathbb{Z}_2$ invariants for 3D topological insulators \cite{Fu2007, Hasan2011}, that is, $B_s = (B_{s_x}^{yz},B_{s_y}^{xz},B_{s_z}^{xy}) $, where $B_{s_k}^{ij}$ is the spin Bott index using the spin operator $\hat{s}_k = \frac{\hbar}{2} \sigma_k$ and $X_i$, $X_j$ position matrices, with $i,j,k = \{x,y,z\}$ for constructing the spin operator projected onto occupied states, Eq. \eqref{eq:proj_Pz}, and the projected position operators, Eq. \eqref{eq:U_proj_position_operators} and \eqref{eq:V_proj_position_operators}, respectively. For strong 3D topological insulators all three indices are equal to unity and weak topological insulators have at least one of the three indices equal to zero. Therefore, the spin Bott index vector corresponds to the surface that displays the topological surface states. In the case of an amorphous phase, all surfaces are equivalent (isotropic) and a single index is sufficient to characterize  the topological phase. Moreover, we confirmed the results of all spin Bott index calculations by calculating the $\mathbb{Z}_2$ invariant. As expected, these invariants agree.

We also use the PAO Hamiltonians to study the energy dispersion of the surface states by means of the spectral function $A(k,\omega)$ computed from the imaginary part of the semi-infinite surface Green's function $A(k,\omega) = -\frac{1}{\pi} \lim_{\eta\rightarrow 0^+} {\rm Im}\,{\rm Tr}\; G_s(k,\omega+i\eta)$. The Green's function $G_s$ of the top and bottom surfaces are obtained recursively by the \citet{Sancho1985} algorithm as implemented in the WannierTools package \cite{WannierTools}.

\section{Results}
\label{sec:results}

\subsection{Amorphous Bi$_2$Se$_3$}

\subsubsection{Structural properties}
\begin{figure}[!htb]
\centering
\includegraphics[width=\linewidth]{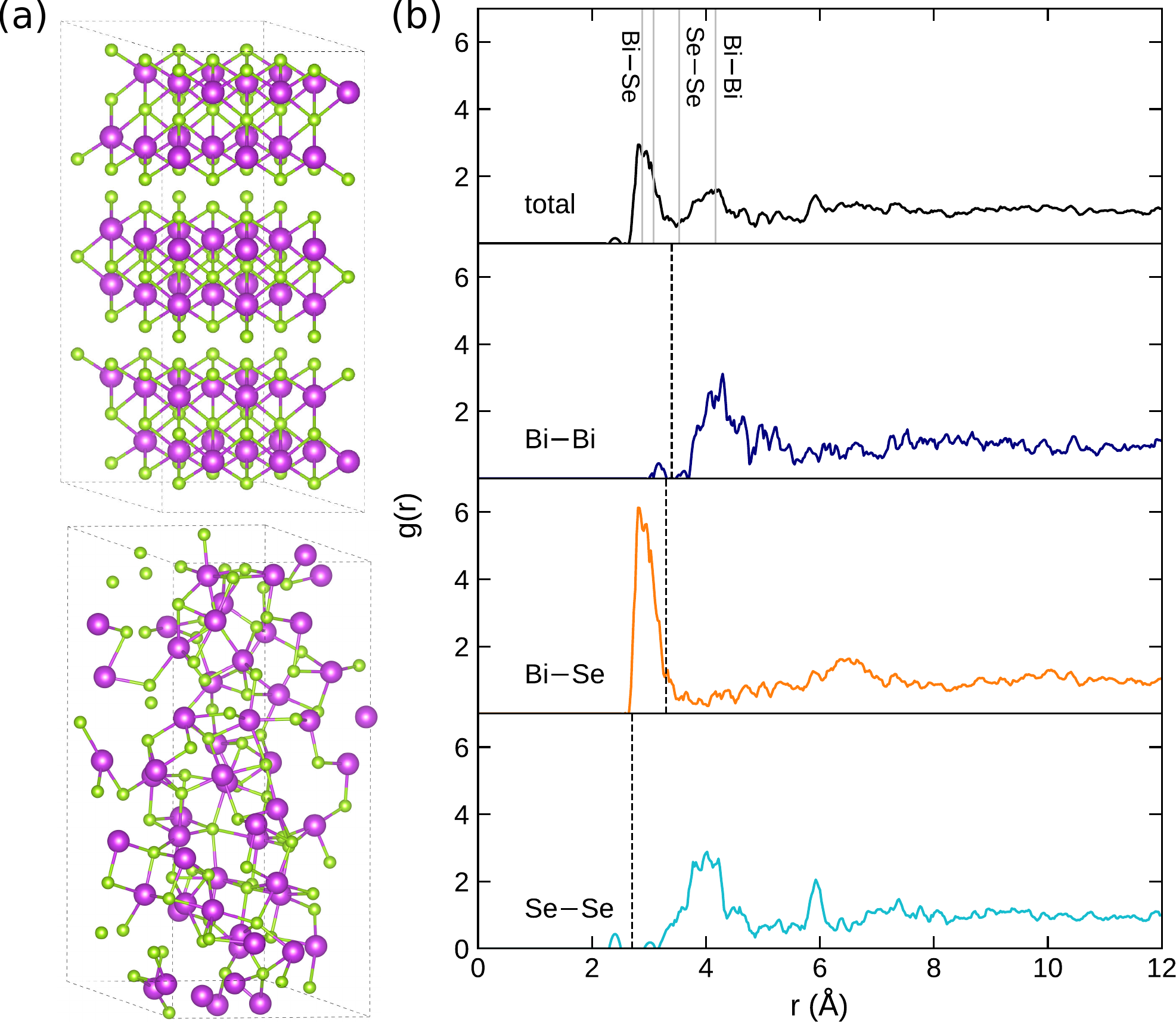}
\caption{Structure and partial pair distribution function of $a$-$\rm Bi_2Se_3$. (a) Structure of crystalline (top) and amorphous (bottom) $\rm Bi_2Se_3$ showing Bi$-$Se bonds. (b) Total and partial pair distribution function (PDF) $g(r)$ of $a$-$\rm Bi_2Se_3$. The first peak of the partial PDF is located at $\SI{3.20}{\angstrom}$ for Bi$-$Bi, $\SI{2.88}{\angstrom}$ for Bi$-$Se, and $\SI{2.41}{\angstrom}$ for Se$-$Se. The second peaks for Bi$-$Bi and Se$-$Se are located at $\SI{4.12}{\angstrom}$ and $\SI{4.05}{\angstrom}$, respectively. The dashed vertical lines mark the bond cutoff. The vertical lines in the total PDF corresponds to the peak position of the partial PDF of crystalline $\rm Bi_2Se_3$.}
\label{fig:structure_partial-pdf}
\end{figure}

The well known $\rm Bi_2Se_3$ crystalline phase belongs to the rhombohedral crystal structure, with space group $R\bar{3}m$, and is composed of inversion symmetric quintuple layers (QLs) with alternating Se and Bi atoms forming a triangular lattice, see Fig. \ref{fig:structure_partial-pdf}(a). In the crystalline phase, the interaction between QLs is of van der Waals type. Within each QL, the Bi atoms are surrounded by Se atoms forming octahedral local environments \cite{Zhang2009}.

We start by analyzing the atomic structures obtained for the amorphous ${\rm Bi}_2{\rm Se}_3$ ($a$-${\rm Bi}_2{\rm Se}_3$). Figure \ref{fig:structure_partial-pdf}(a) shows both the pristine structure of ${\rm Bi}_2{\rm Se}_3$ and an $a$-${\rm Bi}_2{\rm Se}_3$ atomic structure obtained after the melt-quench-anneal procedure. The latter shows no evidence of the quintuple layers found in the crystalline phase. In Fig. \ref{fig:structure_partial-pdf}(b) we report the total and partial pair distribution function (PDF) of $a$-${\rm Bi}_2{\rm Se}_3$. As expected for an amorphous phase, the PDF converges to unity for large distances, losing peak coherence, and indicating long-range disorder \cite{ePDF,ePDF_disorder}. 
For small $r$, we observe two defined peaks for the first and second-neighbor pairs. In the amorphous phase we obtain a small number of homopolar Bi$-$Bi and Se$-$Se bonds, as evidenced by the partial PDFs of Fig.~\ref{fig:structure_partial-pdf}(b), and a majority of Bi$-$Se bonds. Although the partial PDF of Se$-$Se bonds shows a small peak at $\approx\SI{2.2}{\angstrom}$, the first well defined peak is attributed to Bi$-$Se pairs at $\SI{2.88}{\angstrom}$, and the second peak to longer Bi$-$Bi and Se$-$Se bonds near $\SI{4}{\angstrom}$. Comparing with the crystalline peaks, marked by gray lines in the total PDF, the Bi$-$Se bonds show no significant shift from the crystalline values. For the crystalline phase, the Se$-$Se pairs occur between different quintuple layers, while for $a$-${\rm Bi}_2{\rm Se}_3$ we find a small number of Se$-$Se bonds near $\SI{2.41}{\angstrom}$ and a second-neighbor peak at larger distances, $\SI{4.05}{\angstrom}$. For Bi$-$Bi bonds we observe a similar scenario, with the bonds being between second-neighbors within each QLs in the crystalline phase, however in $a$-${\rm Bi}_2{\rm Se}_3$ the calculation gives a small number of homopolar bonds at $\SI{3.20}{\angstrom}$ and a stronger second peak located at $\SI{4.12}{\angstrom}$.

\begin{figure*}[!htb]
\centering
\includegraphics[width=\linewidth]{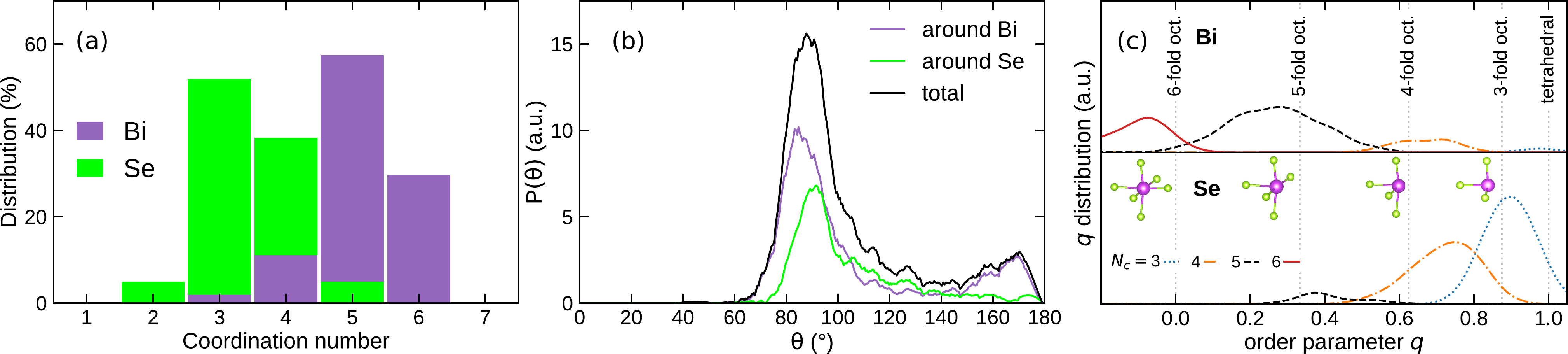}
\caption{(a) Coordination number of Bi and Se atoms in $a$-$\rm Bi_2Se_3$ considering the bond cutoff marked in Fig. \ref{fig:structure_partial-pdf}(b). (b) Angle distribution and (c) distribution of the local order parameter $q$ for $a$-$\rm Bi_2Se_3$ as defined in Eq. \eqref{eq:order_param}. The vertical dashed lines corresponds to $q$ values for ideal geometries. The insets corresponds to the local environments.}
\label{fig:coord_angle_order_param}
\end{figure*}

The distribution of coordination numbers around the Bi and Se atoms is shown in Fig. \ref{fig:coord_angle_order_param}. In the crystalline phases there are only sixfold coordinated Bi atoms while threefold ($33.3\;\%$) and sixfold coordinated ($66.7\;\%$) Se atoms. In the amorphous phase we find a larger number of fivefold coordinated Bi atoms ($57.4\;\%$), indicating a defective local environment for Bi atoms. We also observe a larger number of threefold coordinated Se atoms ($51.8\;\%$), while there is negligible number of sixfold coordinated Se atoms. The average coordination number for each species is presented in Table \ref{tab:avrg_coord_num}. The small number of homopolar bonds is present in the form of dimers (see Fig. \ref{fig:structure_partial-pdf}). This picture is very different from that of $a$-${\rm Sb}_2{\rm Te}_3$ and $a$-GST \cite{Caravati2007,Caravati2010}, where Sb atoms are fourfold coordinated. However, like $a$-${\rm Sb}_2{\rm Te}_3$, chalcogen atoms are mostly threefold coordinated. In addition, Table \ref{tab:avrg_coord_num} and Fig.~\ref{fig:structure_partial-pdf}(b), indicate that the Bi atoms are preferentially bonded to the Se ones.

To investigate the local environment of the Bi and Se atoms, we calculate the distribution of bond angles $P(\theta)$ around each species. Figure \ref{fig:coord_angle_order_param}(b) shows $P(\theta)$ for $a$-${\rm Bi}_2{\rm Se}_3$ obtained from a representative realization of Fig. \ref{fig:structure_partial-pdf}. We find two well defined peaks at ${\sim}\SI{90}{\degree}$ and ${\sim}\SI{170}{\degree}$, around Bi atoms,  which are reminiscent of the octahedral local environment. For Se we only find a single peak at ${\sim}\SI{90}{\degree}$, in contrast to the peaks at $\SI{90}{\degree}$ and $\SI{180}{\degree}$ characteristic of the crystalline case. This is an evidence of defective octahedral environments in $a$-$\rm Bi_2Se_3$. The angle distribution shows that the local environments resembles that of the crystalline phase, however with the lack of sixfold coordinated atoms (see Fig. \ref{fig:coord_angle_order_param}) ruled by the Bi and Se local environments neighboring vacancies. This is similar to $a$-${\rm Sb}_2{\rm Te}_3$ and $a$-GST, where the Sb and Te atoms form defective octahedral environments with different coordinations \cite{Caravati2007,Caravati2010}. In $a$-${\rm Sb}_2{\rm Te}_3$, the Sb atoms are shown to form fourfold coordinated defective octahedra \cite{Caravati2010}. In a similar system, $a$-${\rm Bi}_2{\rm Te}_3$, Bi atoms mostly forms sixfold coordinated octahedra \cite{Guo2019} with a smaller octahedral distortion than in $a$-${\rm Sb}_2{\rm Te}_3$.

To gain further insight on the local geometry of $a$-$\rm Bi_2Se_3$, we compute the distribution of local order parameter $q$, originally introduced to test for tetrahedral configurations \cite{CHAU1998},  defined as \cite{Errington2001}
\begin{equation}
    q_j = 1 - \frac{3}{8} \sum_{i>k} \left(\frac{1}{3} + \cos \theta_{ijk} \right)^2 \;\text{,}\label{eq:order_param}
\end{equation}
where the sum runs over all $i$ and $k$ atoms bonded to the central atom $j$.  Equation \eqref{eq:order_param} gives $q=0$ for a perfect octahedral environment and $q=1$ for perfect tetrahedrals. For ideal defective local environments the order parameter becomes $q=1/3$ for fivefold coordinated octahedra, $q=5/8$ for fourfold coordinated octahedra and $q=7/8$ for threefold coordinated octahedra.

In Fig. \ref{fig:coord_angle_order_param}(c) we show the distribution of $q$ for both the Bi and the Se environments in $a$-${\rm Bi}_2{\rm Se}_3$. Figure \ref{fig:coord_angle_order_param}(c) shows that there are no significant occurrences of tetrahedral environments. For Bi atoms there is a larger distribution of fivefold coordinated octahedra, while Se atoms present threefold coordinated octahedra, followed by fourfold coordinated octahedra, in agreement with the coordination number and angle distribution.

\begin{ruledtabular}
\begin{table}[!htb]
    \centering
    \caption{Average coordination number for each species pair of $a$-$\rm Bi_2Se_3$.}
    \label{tab:avrg_coord_num}
    \begin{tabular}{lccccc}
           & With Bi & & With Se & & Total \\ \hline
        Bi & 0.111   & & 5.278   & & 5.389 \\
        Se & 3.518   & & 0.074   & & 3.592 
    \end{tabular}
\end{table}
\end{ruledtabular}

\subsubsection{Electronic properties}
\begin{figure}[!htb]
\centering
\includegraphics[width=\linewidth]{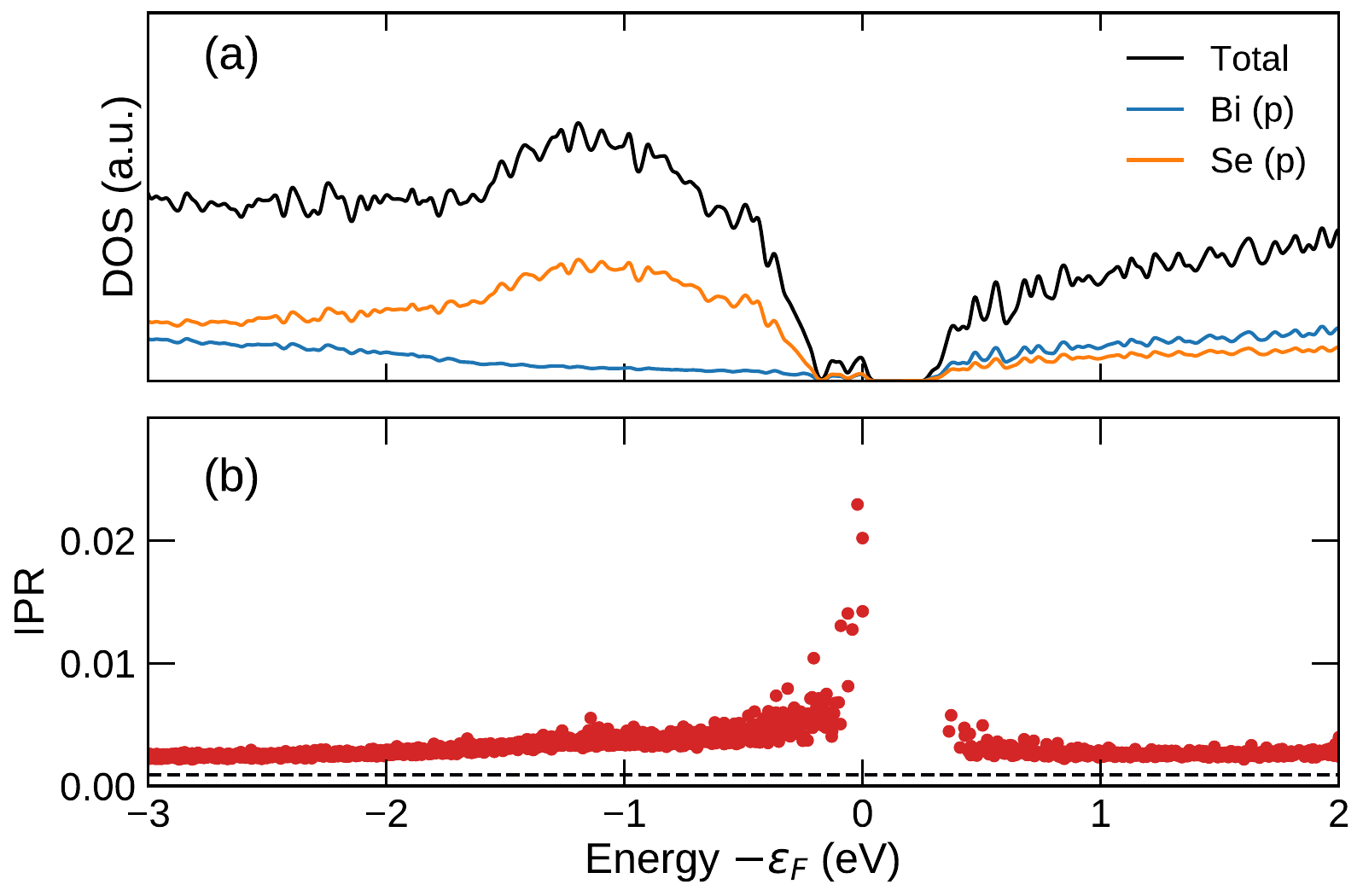}
\caption{(a) Total and partial density of states (DOS) of the representative realization of amorphous $\rm Bi_2Se_3$ system in Fig. \ref{fig:structure_partial-pdf}(a). (b) Inverse participation ratio (IPR) of the same system for k-points inside the BZ sampled with a $3\times 3\times 1$ $\Gamma$-centered Monkhorst-Pack grid. The dashed line corresponds to the limit of maximal delocalization, namely, $1/N$.}
\label{fig:pdos_ipr}
\end{figure}

In Fig. \ref{fig:pdos_ipr}(a) we report the electronic density of states (DOS) of $a$-$\rm Bi_2Se_3$. We find an energy gap of \SI{0.27}{\electronvolt} which is very similar to the energy gap of the crystalline phase, namely, \SI{0.32}{\electronvolt}. Also, in Fig.~S6 of the Supplemental Material \cite{SuppMat} we present a comparison between the DOS obtained with standard PBE for exchange and correlation functional and with the screened hybrid functional HSE06~\cite{Heyd2006}, showing a larger energy gap, namely, \SI{0.58}{\electronvolt}. In the partial DOS of Fig.~\ref{fig:pdos_ipr}(a), the valence band is dominated by Se-\textit{p} orbitals while the conduction band shows an increase in Bi-\textit{p} orbitals. The crystalline phase has an inverted orbital character between Bi and Se $p$ orbitals at the $\Gamma$-point \cite{Liu2010}, which is absent in the amorphous phase. We compute the spin Bott index for $a$-$\rm Bi_2Se_3$ and find $B_s = 0$, that corresponds to a trivial phase. These results are equivalent for all different amorphous structures we have analyzed \cite{SuppMat}.

We characterize the localization properties of $a$-$\rm Bi_2Se_3$ by studying the inverse participation ratio (IPR), namely,
\begin{equation}
    {\rm IPR}_{n,k} = \dfrac{\sum_i^N |\langle i | \psi_{n,k} \rangle |^4}{ \left(\sum_i^N |\langle i | \psi_{n,k} \rangle |^2 \right)^2} \;\text{,} \label{eq:ipr}
\end{equation}
where $\psi_{n,k}$ is the $n$th eigenstate of the local PAO basis Hamiltonian obtained by projecting the KS eigenstates onto the Bi and Se local orbitals $i$, with $N$ the total number of basis orbitals in the cell. We recall that ${\rm IPR} = 1/N$ corresponds to a completely delocalized state, while ${\rm IPR} = 1$ stands for a maximally localized one. Figure \ref{fig:pdos_ipr}(b) shows that the states of $a$-$\rm Bi_2Se_3$ tend to localize near the Fermi level in both valence and conduction bands, while there is dominance of delocalized states at other energies.

\subsubsection{Surface spectra of amorphous Bi$_2$Se$_3$}
\begin{figure}[!htb]
\centering
\includegraphics[width=\linewidth]{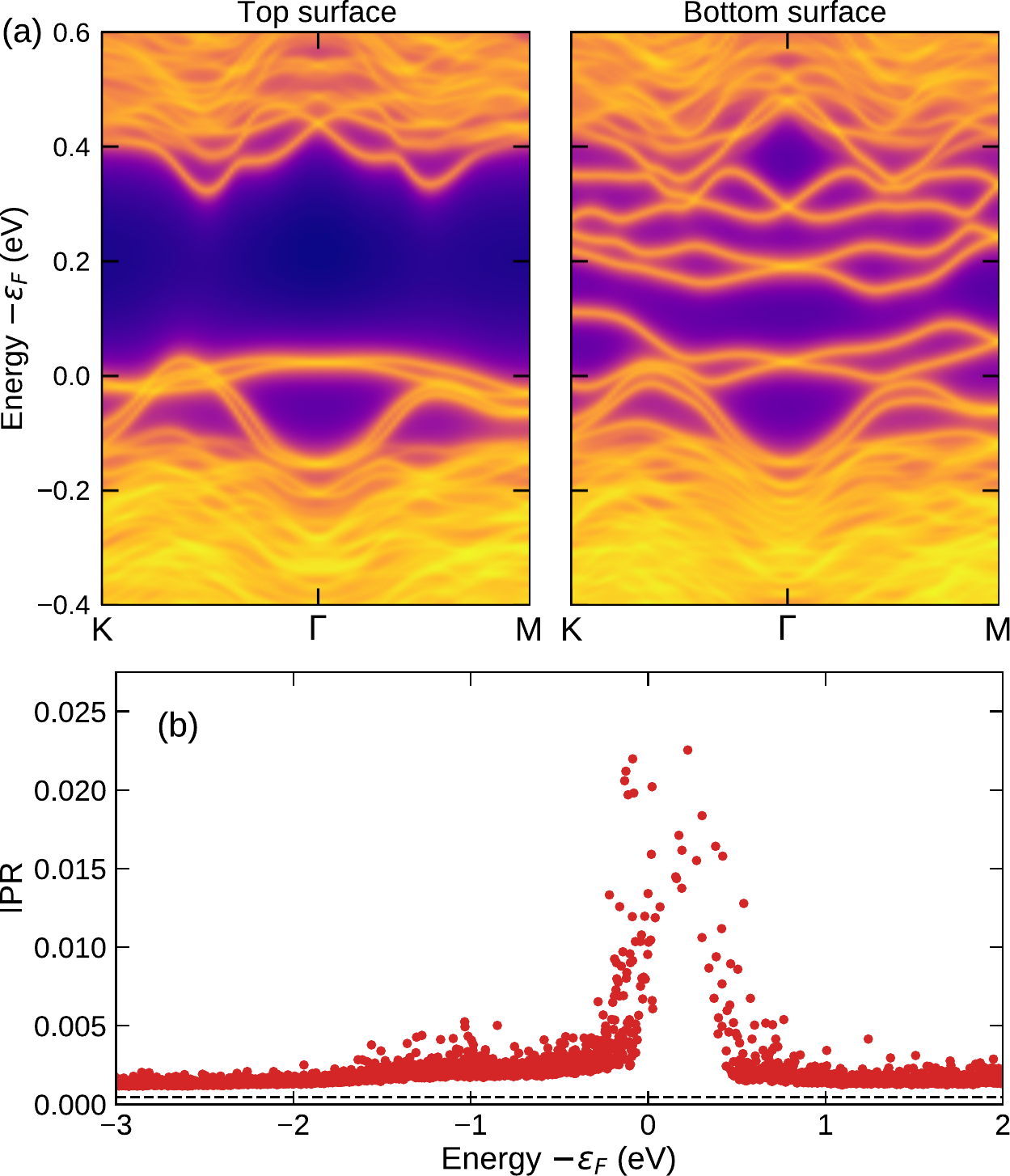}
\caption{(a) Spectral functions computed from the semi-infinite surface Green's function of the top and bottom surfaces. (b) Inverse participation ratio (IPR) of $a$-$\rm Bi_2Se_3$ surface for all k-points in a $3\times 3\times 1$ $\Gamma$-centered Monkhorst-Pack grid. The surface  IPR calculated for a $1\times 1\times 2$ slab. The dashed line corresponds to the limit of 
maximal delocalization, namely, $1/N$.}
\label{fig:surf_states_ipr}
\end{figure}

Figure \ref{fig:surf_states_ipr}(a) presents the electronic spectral function of $a$-$\rm Bi_2Se_3$ computed from the semi-infinite surface Green's function. Given the amorphous nature of the structure, the electronic states at the top and bottom surfaces have very different energy dispersions. In agreement with the computed trivial insulating invariant, there are no states crossing the energy gap connecting conduction and valence bands. Interestingly, we find few low energy dispersion states close to the Fermi level. These low dispersive states of the top (at $\varepsilon_F$) and bottom (up to $0.35$\,eV above $\varepsilon_F$) surfaces have a localized nature as shown by the IPR,  Fig. \ref{fig:surf_states_ipr}(b).

\subsection{Topological phase transition}

The simulations reported above perform the melt-quench-anneal procedure described in Sec.~\ref{sec:methods} on pristine $\rm Bi_2Se_3$ crystalline structures. By heating the systems to $2000\;\rm K$, we find that all analyzed structures end up at a metastable state of the amorphous phase with a topological trivial Bott index, namely, $B_s=0$. In this subsection we study both the structural crystal-amorphous and electronic trivial-topological transitions in more detail.

\subsubsection{Temperature induced topological phase transition}
Let us first  investigating the dependence of the transitions on temperatures ranging form $300$ K up to $2000$ K (for a fixed annealing time).

More specifically, we perform a $10\rm\;ps$ annealing for each considered temperature in the NVT ensemble. The structures are sampled at the end of each annealing. Our simulations show that all structures obtained for annealing temperatures from \SI{300}{\kelvin} up to \SI{1500}{\kelvin} return to the crystalline phase after submitted to a structure relaxation. Only the structures obtained by annealing at \SI{1600}{\kelvin}, \SI{1700}{\kelvin}, and \SI{2000}{\kelvin} remain at the disordered states.

\begin{figure}[!htb]
\centering
\includegraphics[width=\linewidth]{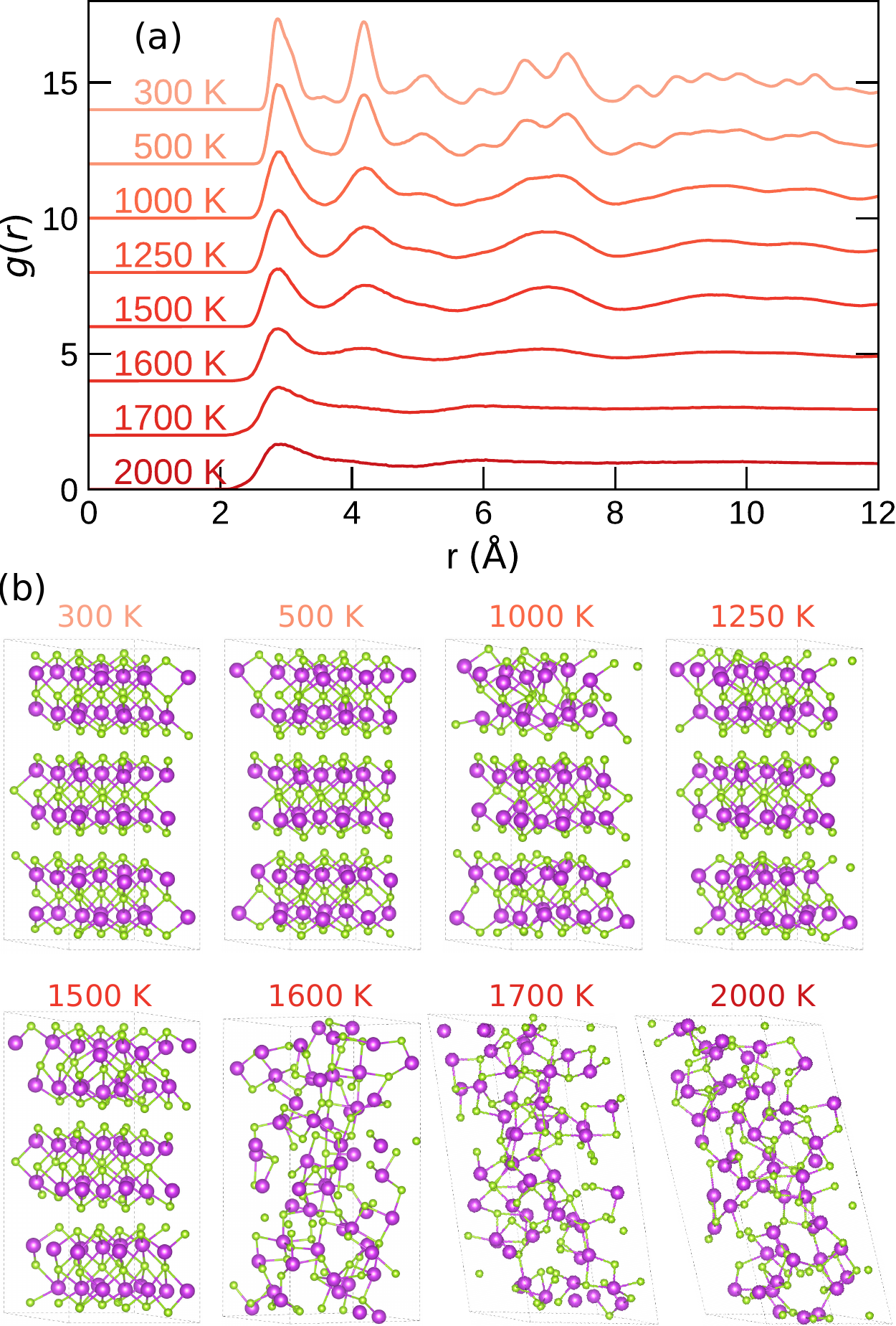}
\caption{(a) Pair distribution function (PDF) of $\rm Bi_2Se_3$ after annealing at the corresponding temperature and (b) structures sampled from the annealing at increasingly high temperatures. In (a) the PDF was obtained by using the time average from each annealing trajectory. In (b) the structures from \SI{1600}{\kelvin} to \SI{2000}{\kelvin} were relaxed.}
\label{fig:transition_pdf_melt_structures}
\end{figure}

Figure \ref{fig:transition_pdf_melt_structures} shows the total PDF and the corresponding sampled geometry obtained for a selection of representative temperatures. At $300\;\rm K$, due to Debye-Waller broadening, the atoms are slightly displaced from their zero-temperature crystalline positions, the quintuple layers are still well defined, and the PDF shows broadened peaks at the corresponding neighbor distances. 
As the temperature increases the PDF peaks lose coherence and only the one corresponding to the first neighbors survive. At $1500\;\rm K$ there is a broad peak between $\SI{6}{\angstrom}$ and $\SI{8}{\angstrom}$, a distance much larger than that between first nearest-neighbor, which disappears at $1600\;\rm K$. 
For $T=1600\;\rm K$, already at the early annealing stages, the studied structures (not shown here) show defective quintuple layers with Bi atoms migrating from the quintuple layer to the region between two quintuple layers. As the annealing proceeds, the crystalline-amorphous transition takes place, and at the end, the structures are completely distorted, as shown in Fig. \ref{fig:transition_pdf_melt_structures}(b). 

\begin{figure}[!htb]
\centering
\includegraphics[width=\linewidth]{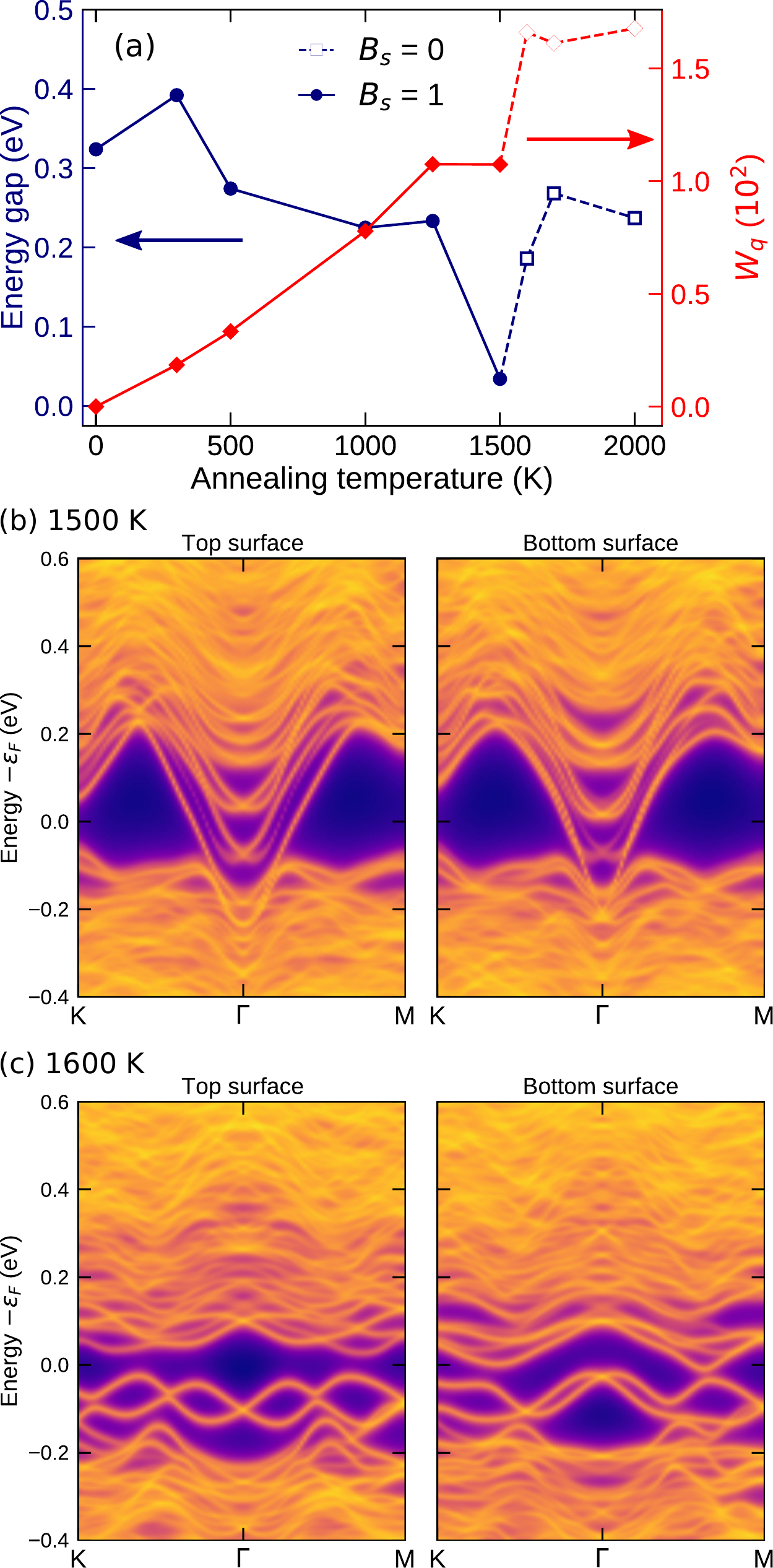}
\caption{(a) Energy gap and disorder $W_q$ for structures sampled from the annealing at increasingly high-temperatures. The structures from \SI{1600}{\kelvin} to \SI{2000}{\kelvin} were relaxed. The disorder is measured by the sum of Bi local order parameter for three, four, and fivefold coordinated environments divided by the number of Bi atoms, $W_q = \sum_j q_j\, / N_{\rm Bi}$. (b) and (c) spectral functions computed from the semi-infinite surface Green's function of the top and bottom surfaces for the structures in Fig. \ref{fig:transition_pdf_melt_structures}(b) annealed at (b) \SI{1500}{\kelvin} and at (c) \SI{1600}{\kelvin}.}
\label{fig:transition_disorder_surf_states}
\end{figure}

To investigate the topological phase transition we compute the energy gap and the spin Bott index as a function of the annealing temperature. To characterize the crystal-amorphous transition we introduce a global disorder measure, defined as the sum of the Bi local order parameter for three, four, and fivefold coordinated environments divided by the number of Bi atoms, $W_q = \sum_j q_j\, / N_{\rm Bi}$. 
This measure of disorder captures very nicely the disorder resulting from defective octahedral environments. The results are shown in Fig. \ref{fig:transition_disorder_surf_states}(a). As the annealing temperature increases, the energy gap decreases as a consequence of increasing disorder. At $1500 \;\rm K$ the energy gap reaches a minimum and increases again as the temperature grows further. In this case, the contribution of defective regions to the states near the Fermi level is increased. Note that the pronounced structural distortion for $T \gtrsim 1500\;\rm K$, see Fig.~\ref{fig:transition_pdf_melt_structures}, is accompanied by a topological-trivial phase transition characterized by the behavior of the spin Bott index, as shown in Fig. \ref{fig:transition_disorder_surf_states}(a). These results strongly support that the defective octahedra environments rule a topological transition in $a$-$\rm Bi_2Se_3$. We find a similar behavior in other bismuth-selenide crystalline phases (see discussion of Section IV of the Supplemental Material ~\cite{SuppMat}). To evaluate the robustness of the transition, we have arbitrarily increased the SOC strength up to $1.5$ times the \textit{ab initio} value. Within this range, we find that the system always remains trivial for temperatures above the $T = 1500\rm\; K$.

Figures \ref{fig:transition_disorder_surf_states}(b) and (c) show ${\rm Bi}_2{\rm Se}_3$ surface spectra for temperatures near the topological-trivial phase transition point. At $1500\rm\;K$, within the topological phase ($B_s = 1$), both surfaces have similar spectra characterized by surface states connecting the valence and conduction bands. Also, these surface states overlap with the valence bands for energies around $-0.2\;\rm eV$. At $1600\rm\;K$, the system is already topologically trivial ($B_s=0$) and the surface states spectrum display an energy gap. Interestingly, Fig.~\ref{fig:transition_disorder_surf_states}(c) shows disorder-induced states localized at the surfaces inside the bulk energy gap. The states with energies between $-0.1\rm\; eV$ and $0.05\rm\; eV$ have a larger contribution from the surface orbitals.

\subsubsection{Disorder induced topological transition}
To increase the disorder in a controllable manner and to measure the direct effect over the surface states, we place a crystalline structure in contact with a $2000\;\rm K$ heat bath in an NVT ensemble and sample structures at the beginning of the annealing processes as a function of time, at every 10 fs. We compute the energy gap, spin Bott index, disorder $W_q$ and surface spectra as a function of annealing time.

\begin{figure}[!htb]
\centering
\includegraphics[width=\linewidth]{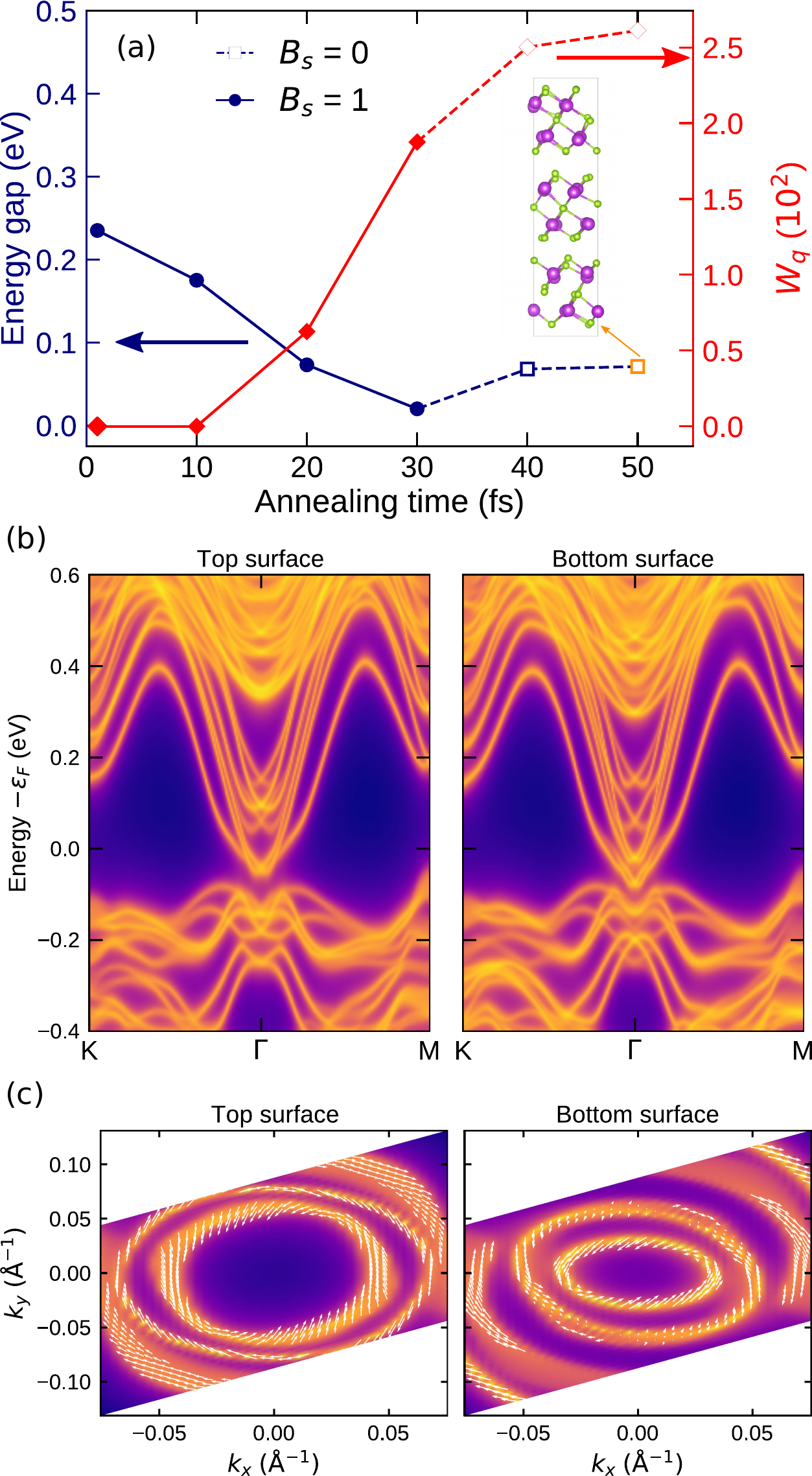}
\caption{(a) Energy gap and disorder $W_q$ for structures sampled from annealing the crystalline structure at \SI{2000}{\kelvin}. The disorder is measured by the sum of Bi local order parameter for three, four, and fivefold coordinated environments divided by the number of Bi atoms, $W_q = \sum_j q_j\; / N_{\rm Bi}$. (b) Spectral functions computed from the surface Green's function of the top and bottom surfaces for the structure at $\rm 50\;fs$ marked in orange in (a). (c) Spin-texture of top and bottom surface spectral functions at $\varepsilon_F$, $0.2$ around the $\Gamma$-point, in fractional units. In (a) the inset corresponds to the structure obtained in $50\;\rm fs$.}
\label{fig:const_temp_transition}
\end{figure}

Figure~\ref{fig:const_temp_transition}(a) shows that the disorder increases monotonically with annealing time while the energy gap decreases almost linearly until reaching values close to zero. At $30\;\rm fs$ the energy gap is $20\;\rm meV$ and the structure still displays a topological phase ($B_s = 1$). After $30\;\rm fs$ the energy gap increases and the system electronic structure is tolologically trivial ($B_s = 0$). 
The surface spectrum in Fig. \ref{fig:const_temp_transition}(b) at $50\;\rm fs$ corresponds to the trivial phase. Interestingly, we still observe a reminiscent of the topological phase. The surface states are gapped and trivial. In Fig.~\ref{fig:const_temp_transition}(c), the surface states show a Rashba-like spin texture. The spin orientation is approximately perpendicular to the k-vector while also breaking spin degeneracy away from the $\Gamma$-point. Unfortunately, the small energy gap in the surface spectrum is inaccessible to experimental techniques.  
Although trivial, the surface spectrum of Fig. \ref{fig:const_temp_transition}(b) is similar to the one for $1500\;\rm K$ in Fig.~\ref{fig:transition_disorder_surf_states}(b) that corresponds to a disordered structure with a helical surface Dirac cone. 

For $T=2000$ K, both the crystalline-amorphous and the topological-trivial transitions take place already at very short annealing times $t_{\rm anneal} \gtrsim 30$ fs. 
We note that by quenching to \SI{300}{\kelvin} the temperature of the structures obtained for $t_{\rm anneal} = 40$ and $50\;\rm fs$ they converge to a glassy state instead of returning to the crystalline phase. This is consistent with the picture we have discussed when analyzing Fig. \ref{fig:transition_pdf_melt_structures}(b). These results suggest that for $T \approx 1500\;\rm K$ the systems reach a structural disorder threshold beyond which the glassy phase is energetically more favorable. The structural transition is accompanied by a topological electronic one, that is for $T \gtrsim 1500\;\rm K$ the energy spectrum is changed and resembles the one found by the amorphous phases in Fig. \ref{fig:pdos_ipr}.

Our simulations are at odds with the findings reported by Ref.~\cite{Corbae2019}. It is conceivable that surface reconstruction on the amorphous surface can modify the nature of the trivial surface states. In the Supplemental Material \cite{SuppMat} we explore this idea considering: (i) a crystalline reconstruction, and (ii) a selenium capping layer scenarios. Our simulations do not indicate significantly change of the general picture described above. On (i) highly dispersive trivial states also arise similar to the ones with disorder close to the topological transition while on (ii) the trivial states remain with low dispersion.

In two-dimensional amorphous bismuthene, departing from the crystalline structure, the topological phase is kept throughout the amorphization \cite{Focassio2021}. Despite the  defective areas created by amorphization, the orbital contribution and average coordination (${\sim}2.7$) remains similar to the one in the crystal. In contrast, for $a$-$\rm Bi_2Se_3$ we show in Fig. \ref{fig:pdos_ipr}(a) that the orbital character near the band edges is different from the one in the crystal. This is attributed to the increase of defective local environments with lower coordination as seen in Fig. \ref{fig:transition_disorder_surf_states}(a) and \ref{fig:const_temp_transition}(a). This reveals a lack of robustness of the topological phase in $\rm Bi_2Se_3$. Comparatively, changing the coordination of Bi atoms by ${\sim}10\%$ (i) does not change the topological character of bismuthene while (ii) it turns $\rm Bi_2Se_3$ into a trivial insulator.

\section{Summary and Conclusions}
\label{sec:conclusions}

We have proposed a model to generate and study the amorphous phase of $\rm Bi_2Se_3$ by quenching from the melt through \textit{ab initio} molecular dynamics simulations. Amorphous $\rm Bi_2Se_3$ presents a small number of homopolar bonds,  with the first neighbor peaks of the PDF dominated by Bi$-$Se pairs. The Bi atoms occupy mostly fivefold coordinated defective octahedral environments, while the Se ones are mostly in threefold coordinated environments.

Different from the crystalline phase, the realization of amorphous $\rm Bi_2Se_3$ yields a trivial insulator characterized by a zero spin Bott index. By sampling different structures from annealing simulations, we study the structural phase transition from the crystalline to the amorphous phase and the topological phase transition induced by disorder. Here, we show that the increase in defective octahedral environments drives a topological phase transition. In contrast with the two-dimensional case, where the topological phase is robust and present even in highly amorphous structures, the topological character of amorphous $\rm Bi_2Se_3$ is more fragile to structural disorder.

Additionally, we demonstrate that the surface states close to the topological phase transition show a memory effect, preserving the helical spin texture of the topological phase surface states, even in the trivial phase. 

This work reports the paramount importance of orbital contributions and local atomic environments in determining the electronic structure and topological nature of solids. We provide insightful information for the search of material-specific topological phases in three-dimensional amorphous materials.

\begin{acknowledgments}
The authors acknowledge Paul Corbae for useful discussions. This work is supported by FAPESP (Grants 19/04527-0, 19/20857-0, 17/18139-6, and 17/02317-2), FAPERJ (Grants E-26/2020.882/2018 and E-26/010.101126/2018), CNPq (Grant 313059/2020-9), and INCT. The authors acknowledge the Brazilian Nanotechnology National Laboratory (LNNano/CNPEM, Brazil) and the SDumont supercomputer at the Brazilian National Scientific Computing Laboratory (LNCC) for computational resources.
\end{acknowledgments}

%



\section*{Supplementary material}

# Amorphous $Bi_2Se_3$ structural, electronic, and topological nature from first-principles

Bruno Focassio,[1, 2, *] Gabriel R. Schleder,[1, 2, 3] Felipe Crasto de Lima,[2, 4] Caio Lewenkopf,[5] and Adalberto Fazzio[2, 1, 4, †]

[1]*Federal University of ABC (UFABC), 09210-580 Santo André, São Paulo, Brazil*
[2]*Brazilian Nanotechnology National Laboratory (LNNano),
CNPEM, 13083-970 Campinas, São Paulo, Brazil*
[3]*John A. Paulson School of Engineering and Applied Sciences,
Harvard University, Cambridge, Massachusetts 02138, USA*
[4]*Ilum School of Science, CNPEM, 13083-970 Campinas, São Paulo, Brazil*
[5]*Instituto de Física, Universidade Federal Fluminense, 24210-346 Niterói, Rio de Janeiro, Brazil*

## I. FULLY RELATIVISTIC DFT *VS* PAOFLOW BAND STRUCTURE FOR BULK $Bi_2Se_3$

In Fig. S1 we show a comparison of the band structure calculated with fully-relativistic DFT by the Quantum Espresso code [1, 2] and the one corresponding to the PAO Hamiltonian derived using the PAOFLOW code [3] accounting for the SOC interaction via an effective approximation [4] using $\lambda_{Bi} = 1.615$ eV and $\lambda_{Se} = 0.320$ eV. Fig. S1 shows that the PAO Hamiltonian accurately describes the energy bands and topological character of $Bi_2Se_3$, reducing the computational cost for the subsequent calculations of topological invariant, inverse participation ratio and surface spectral function considering the amorphous systems.

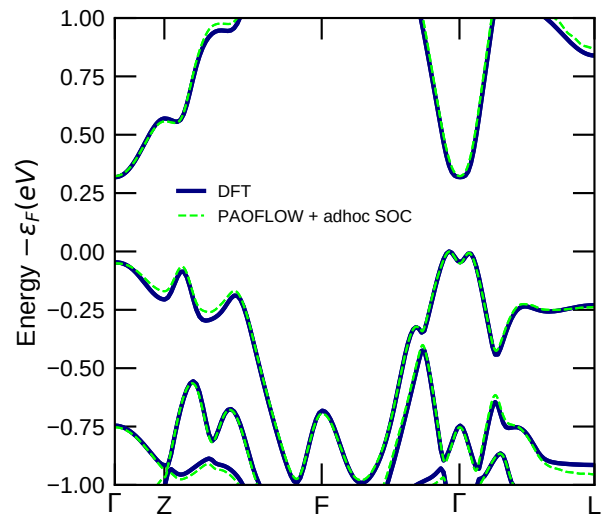

FIG. S1. Comparison of the band structure of the crystalline phase of bulk $Bi_2Se_3$ between the fully relativistic DFT calculation with the Quantum Espresso and the one computed with PAO Hamiltonian via the PAOFLOW code with effective approximation for SOC.

## II. STRUCTURAL CHARACTERIZATION OF OTHER AMORPHOUS REALIZATIONS

In Figs. S2, S3 and S4 we present the structural characterization of different amorphous realizations, named hereafter as #2, #3, and #4, with #1 the one presented in the main text. The realizations #2 and #3 were obtained using the same procedure highlighted in the main text used to generate #1. As stated in the methods section, to

---

* b.focassio@ufabc.edu.br
† adalberto.fazzio@lnnano.cnpem.br

obtain different structure following the same procedure, the melting step is extended by 2 ps for each new structure, after melting the quench, anneal and relaxation procedures are performed with 10 ps for each step. The structure #4 is obtained also using the same procedure, but considering an NPT ensemble were the pressure is fixed at 0 kbar by using a Langevin dynamics [5] combined with Parrinello-Rahman [6, 7] barostat. Figures Fig. S2 to S4 indicate that all obtained geometries for $a$-$Bi_2Se_3$ are structurally equivalent, supporting that the conclusions draw in the main text are robust and general, regardless of the structure considered.

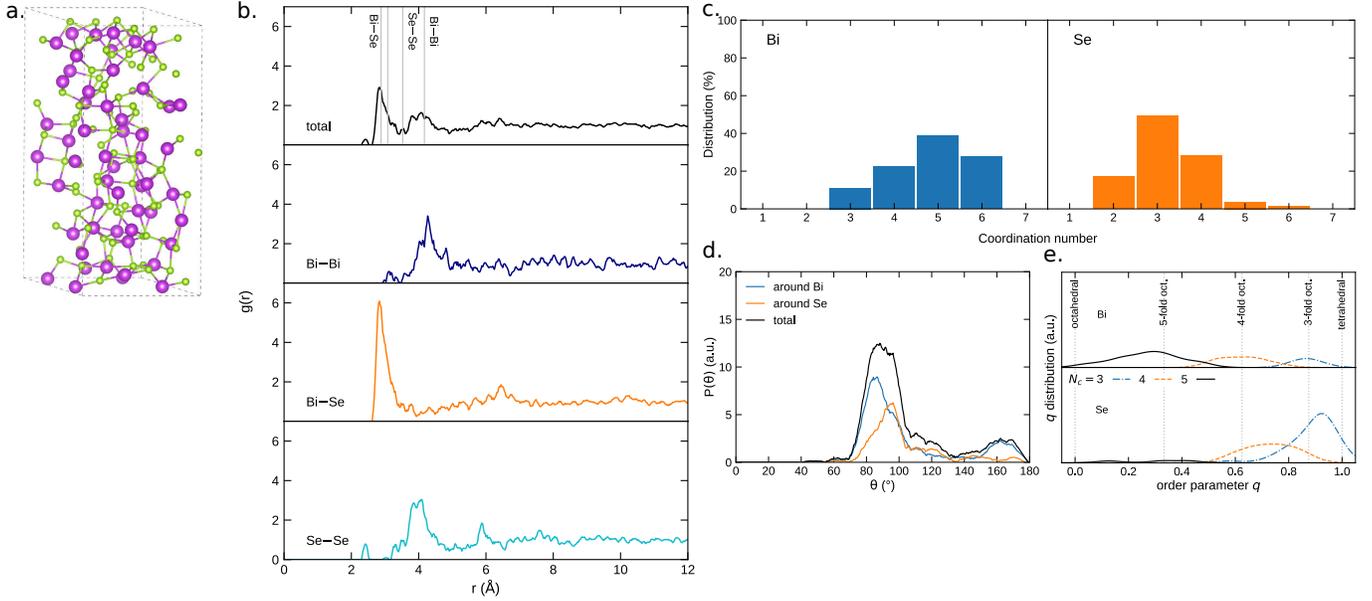

FIG. S2. Structure characterization of $a$-$Bi_2Se_3$ #2. (a) Structure of amorphous $Bi_2Se_3$ showing the Bi−Se bonds. (b) Total and partial pair distribution function (PDF) of $a$-$Bi_2Se_3$. The vertical lines in the total PDF correspond to the peak position of the partial PDF of crystalline $Bi_2Se_3$. (c) Coordination number distribution of Bi and Se atoms in $a$-$Bi_2Se_3$. (d) Angle distribution $P(\theta)$ and (e) local order parameter $q$ distribution, the vertical lines correspond to $q$ for ideal geometries.

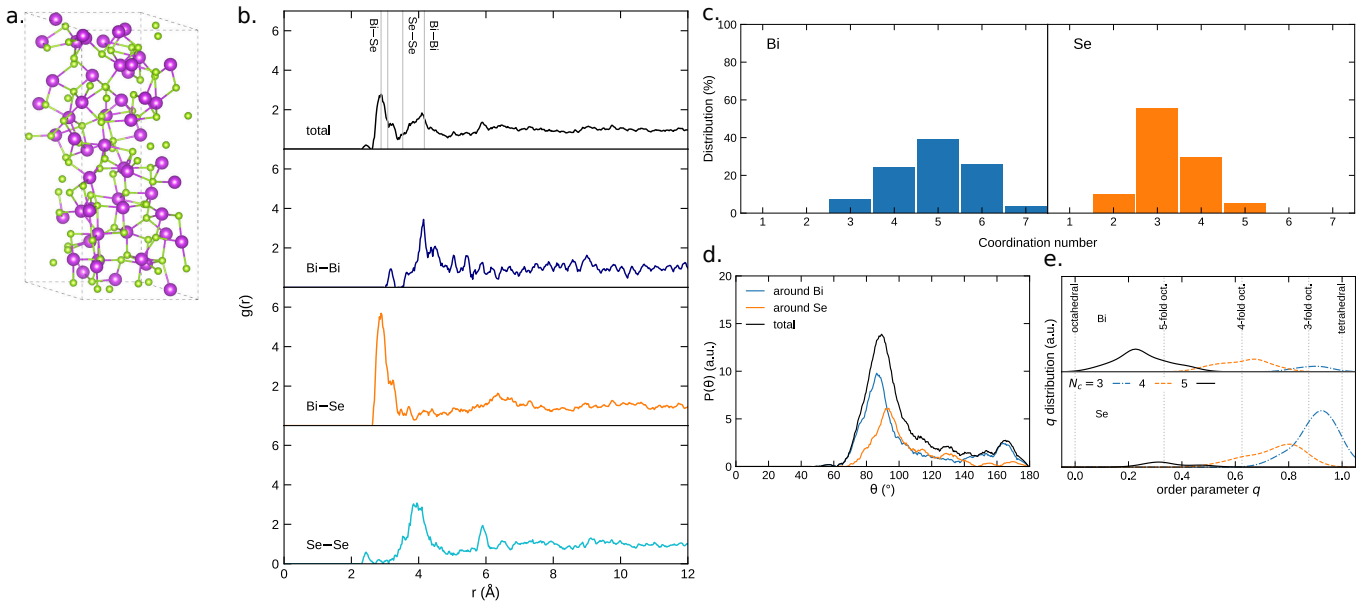

FIG. S3. Structure characterization of $a$-$Bi_2Se_3$ #3. (a) Structure of amorphous $Bi_2Se_3$ showing the Bi−Se bonds. (b) Total and partial pair distribution function (PDF) of $a$-$Bi_2Se_3$. The vertical lines in the total PDF correspond to the peak position of the partial PDF of crystalline $Bi_2Se_3$. (c) Coordination number distribution of Bi and Se atoms in $a$-$Bi_2Se_3$. (d) Angle distribution $P(\theta)$ and (e) local order parameter $q$ distribution, the vertical lines correspond to $q$ for ideal geometries.

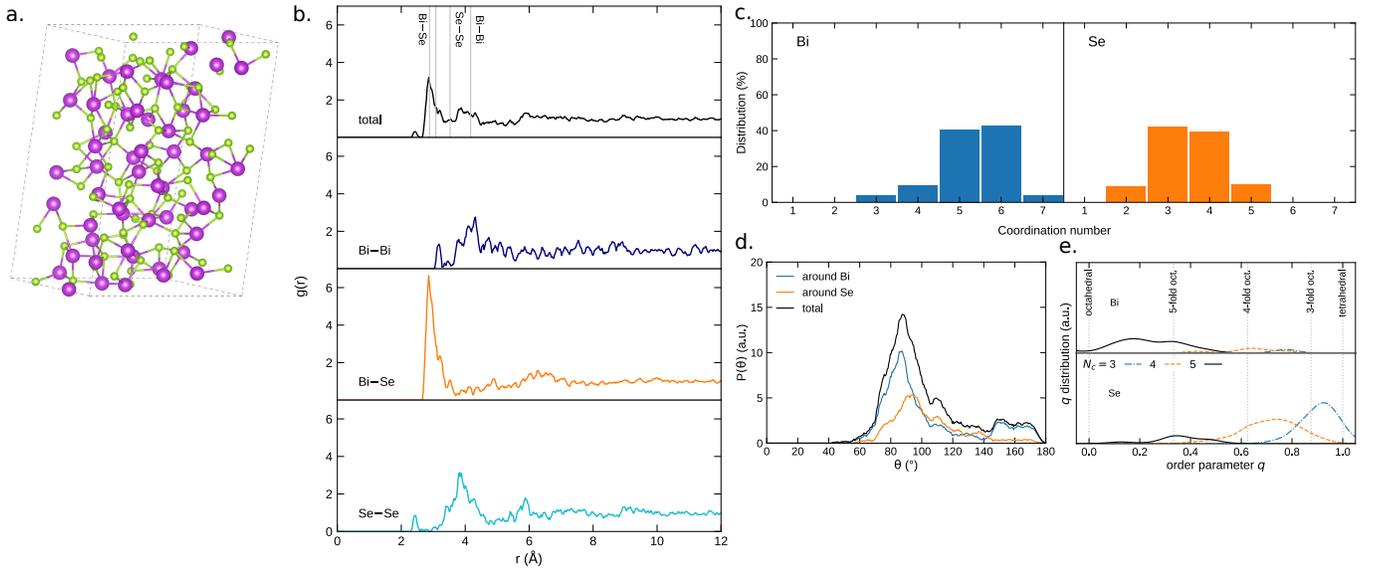

FIG. S4. Structure characterization of $a$-Bi$_2$Se$_3$ #4. (a) Structure of amorphous Bi$_2$Se$_3$ showing the Bi−Se bonds. (b) Total and partial pair distribution function (PDF) of $a$-Bi$_2$Se$_3$. The vertical lines in the total PDF correspond to the peak position of the partial PDF of crystalline Bi$_2$Se$_3$. (c) Coordination number distribution of Bi and Se atoms in $a$-Bi$_2$Se$_3$. (d) Angle distribution $P(\theta)$ and (e) local order parameter $q$ distribution, the vertical lines correspond to $q$ for ideal geometries.

## III. ELECTRONIC STRUCTURE OF OTHER AMORPHOUS REALIZATIONS

Figure S5 presents the partial DOS and inverse participation ratio (IPR) for the amorphous realizations #2, #3, and #4. Again, all amorphous realizations considered have very similar electronic structures. We observe the presence of disorder induced localized states inside the energy gap, mainly in realizations #3 and #4, decreasing the fundamental energy gap.

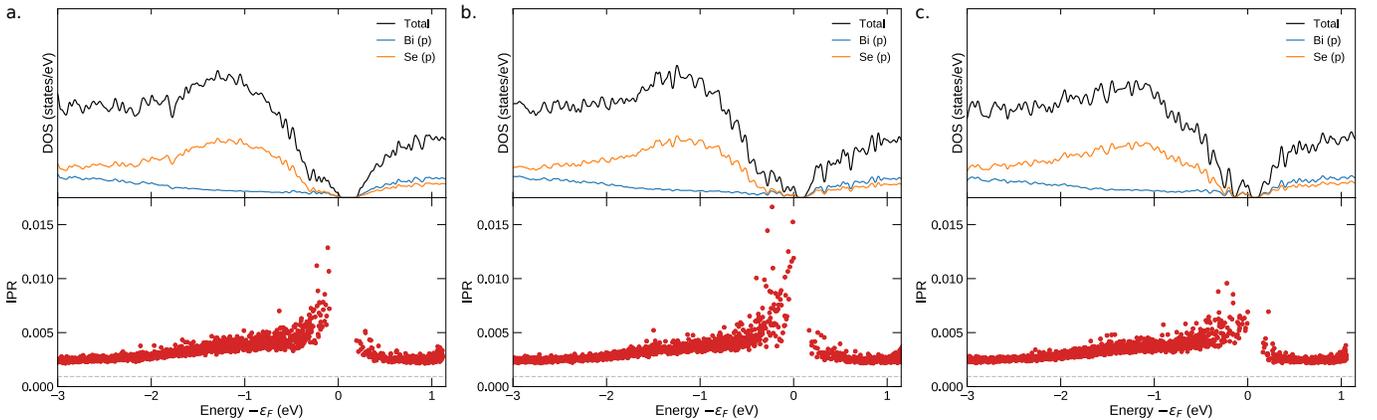

FIG. S5. Total and partial density of states (DOS) of amorphous Bi$_2$Se$_3$ (top) and Inverse participation ratio (IPR) (bottom) for k-points inside the BZ sampled with a $3 \times 3 \times 1$ Γ-centered Monkhorst-Pack grid. The dashed line correspond to the limit of complete delocalization ($1/N$). The panels a, b, and c corresponds to the realization #2, #3, and #4 of $a$-Bi$_2$Se$_3$. All realizations of $a$-Bi$_2$Se$_3$ result in trivial insulator by $B_s = 0$. The energy gap is 0.210 eV, 0.123 eV, and 0.155 eV, respectively.

In Fig. S6 we present a comparison between the electronic density of states obtained with the standard PBE for the exchange and correlation functional with the screened hybrid functional HSE06 [8] for the realization #1 of $a$-Bi$_2$Se$_3$, as discussed in the main text. We find that the energy gap is increased from 0.272 eV with PBE to 0.585 eV with the HSE06 hybrid functional.

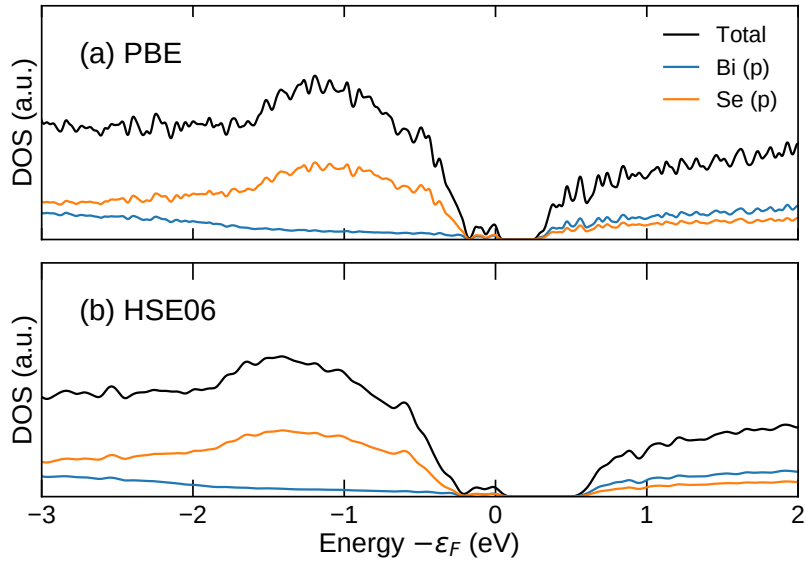

FIG. S6. Electronic total total and partial density of states (DOS) of amorphous $Bi_2Se_3$ #1 with (a) PBE and (b) HSE06. With PBE the energy gap is 0.272 eV, while with HSE06 the energy gap is 0.585 eV.

## IV. IMPACT OF OCTAHEDRA ON TOPOLOGY OF BI,SE-PHASES

We stress the importance of the perfect octahedra local environment (POLE) on the electronic structure topology of the $a$-$Bi_2Se_3$ by considering other bismuth-selenide phases, with and without POLE. We have simulated the orbital projected band structure of the three $Bi_2Se_3$ phases with lower energy-above-hull listed in the Materials Project database [9], and a BiSe phase where all the atoms are in a POLE. Figure S7 shows the crystal and electronic dispersion of these structures. The $Bi_2Se_3$ phase with $P4_2nmc$ space-group has a sixfold coordinated distorted octahedra phase and the Pnma space-group has half of Bi with a distorted sixfold octahedra, and the other half with a fivefold coordinated environment. In these two systems, the band structure presents an insulating character, the Se-$p$ and Bi-$p$ orbitals are not inverted, a typical feature of a trivial phase. In contrast, the $Bi_2Se_3$ R-3m phase, where all bismuth atoms have a POLE, is topological and the states close to the bandgap show the characteristic inversion between Bi and Se $p$-orbitals. phase. Additionally, within the BiSe with Fm-3m space-group, all bismuth atoms have a POLE, that is also translating into an inverted bandgap (non-trivial phase). In the last case the band inversion lies at energies below the Fermi level. However, with an extra Se n-type doping, the correlation between the POLE and the inverted gap can, in principle, be observed.

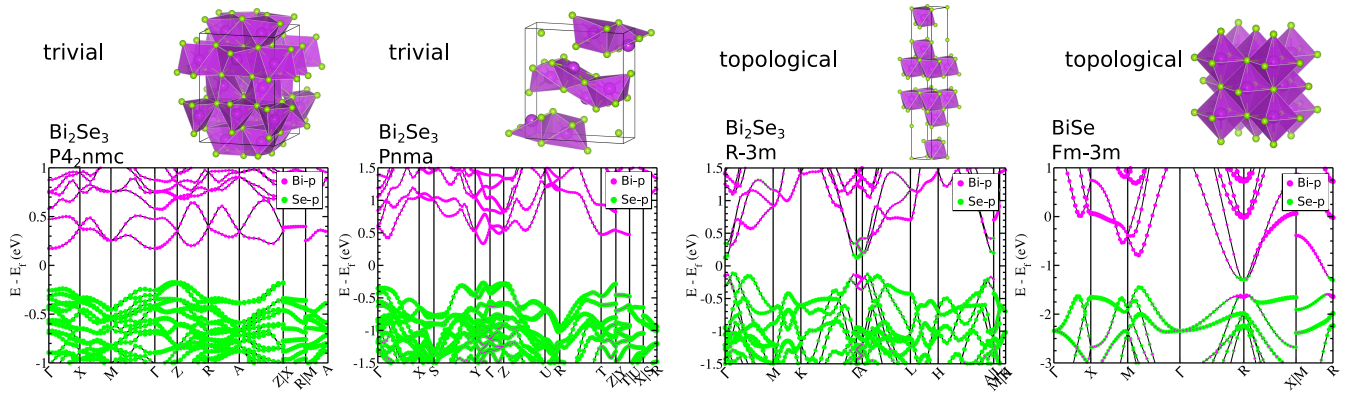

FIG. S7. Different phases of binary (Bi,Se) compounds highlighting the inverted character of the bandgap with particular Bi-Se octahedral structure.

## V. SURFACE RECONSTRUCTION

### A. Crystalline surface reconstruction

We have shown that the $a$-Bi$_2$Se$_3$ is a trivial insulator with trivial and low dispersive surface states. Let us investigate whether surface crystalline reconstructions can lead a different scenario for the surface states energy dispersion. Here, we consider two different possibilities.

First, we study a crystalline Bi$_2$Se$_3$ monolayer covering an amorphous slab, as illustrated by Fig. S8(a). The projection of the bulk and surface states contributions of the slab to the band structure is shown in Fig. S8(b). In contrast to the low dispersive bulk states of the amorphous structure, the pristine surface layer are characterized by well defined highly dispersive states. We note that the energy shift of the lower (S1-pr) and upper (S2-pr) surface is a spurious effect the different amorphous lower and upper terminations in the *finite* slab approximation. Although the surface dispersive states look similar to the pristine Bi$_2$Se$_3$ topological ones, they are not helical and, thus, likely to be originated by a (trivial) Rashba-like dispersion. That is somewhat expected, since both the amorphous structure and the pristine Bi$_2$Se$_3$ monolayer are trivial structures.

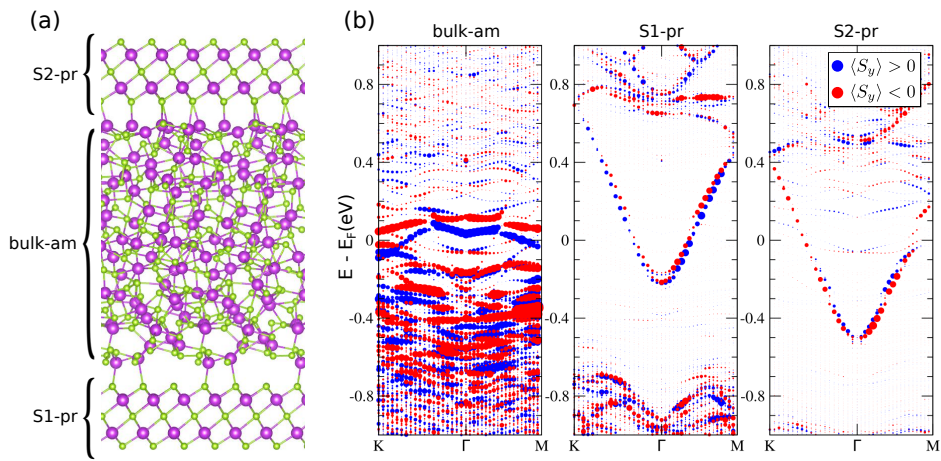

FIG. S8. (a) Atomic structure of the crystalline Bi$_2$Se$_3$ monolayer on top of an amorphous slab. (b) Spin projected contributions on the band structure, e.g. $|\langle \phi_i, s_y | n, k, \sigma \rangle|^2$ summed over the $\phi_i$ atomic orbitals of each region.

### B. Se passivation

Finally, we study surface reconstructions driven by additional Se atoms passivating the surfaces of $a$-Bi$_2$Se$_3$. We consider a $1 \times 1 \times 2$ slab with 20 Å of vacuum. We add 5 Se atoms to the slab geometry. The Se atoms were manually arranged at the surface in order to passivate the dandling bonds. We submit the slab to a 2000 K heat bath (NVT) for 2 ps and anneal it at 300 K for another 10 ps. After the annealing, we identify prominent Se$_2$ groups at the surface, which are removed. The final geometry is relaxed, see Fig. S9(a), and its electronic structure in shown in Fig. S9(b). We observe a number of low dispersive trivial states, similar to the case obtained by the semi-infinite surface energy spectra discussed in the main text. Figure S9(c) also shows the spin texture contribution of atoms belonging to the top and bottom surfaces of the slab, highlighting a small spin splitting, mainly at the bottom surface.

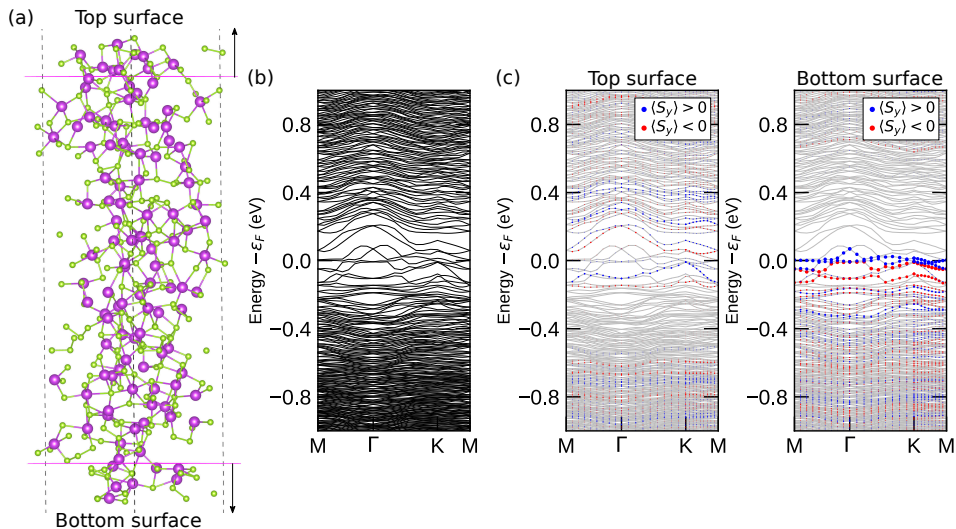

FIG. S9. (a) Atomic structure of the $a$-Bi$_2$Se$_3$ $1 \times 1 \times 2$ slab to study the surface reconstruction. (b) Slab band structure. (c) Spin projected contributions on the band structure, e.g. $|\langle \phi_i, s_y | n, k, \sigma \rangle|^2$ summed over the $\phi_i$ atomic orbitals of each region comprehended by the top and bottom surface atoms demonstrated in panel (a).



\end{document}